\documentclass[%
 preprint,
superscriptaddress,
nolongbibliography,
 amsmath,amssymb,
 aps,showkeys, 
]{revtex4-2}

\usepackage{color}
\usepackage{subcaption}
\usepackage{ulem}
\usepackage{graphicx}
\usepackage{dcolumn}
\usepackage{bm}
\usepackage{hyperref}
\usepackage{braket}
\usepackage[a4paper]{geometry}
\usepackage{tensor}
\usepackage{comment}

\begin{document}


\title{Tomographic entanglement indicators in a coupled oscillator model}
 
\author{Sreelekshmi Pillai}
\email{sreelekshmi@physics.iitm.ac.in}
\author{S. Ramanan}

\affiliation{Department of Physics, Indian Institute of Technology Madras, Chennai, India \looseness=-1}
\affiliation{Center for Quantum Information, Communication and Computing (CQuICC), Indian Institute of Technology Madras, Chennai, India \looseness=-1}
\author{V. Balakrishnan}
\author{S. Lakshmibala}
\affiliation{Center for Quantum Information, Communication and Computing (CQuICC), Indian Institute of Technology Madras, Chennai, India \looseness=-1}
 
\date{\today}

\begin{abstract}
    \noindent We study entanglement in a simple model comprising two coupled linear harmonic oscillators of the same natural frequency. The 
    system is separable in the center of mass (COM) and relative coordinates into two oscillators of frequency $\omega_c$ and $\omega_r$. We 
    compute standard entanglement measures (subsystem linear entropy and subsystem von Neumann entropy) as well as several tomographic 
    entanglement indicators (Bhattacharyya distance, Kullback-Leibler divergence and inverse participation ratio) as functions of the frequency 
    ratio $\eta = \omega_c/\omega_r$, keeping the COM oscillator in its ground state. We demonstrate that, overall, the entanglement indicators reflect quite faithfully the variations in the standard measures. The entanglement is shown to be minimum at $\eta = 1$ and maximum as $\eta \to 0$ or $\infty$.  
\end{abstract}

                              
\maketitle


\section{Introduction}
Entanglement is an important resource for several tasks pertaining to quantum information and computation. Investigations of entanglement have 
been carried out on discrete variable (DV) systems such as two-level atomic ensembles and spin arrays, and on continuous variable (CV) and 
hybrid quantum (HQ) systems as with atoms interacting with a radiation field. These investigations have led to a fair degree of understanding 
of diverse aspects such as the sudden death of entanglement 
during the time evolution of specific systems \cite{yu2009sudden}, the efficacy of different entanglement indicators (see, for instance, \cite{horodecki2009quantum}), signatures of entanglement in avoided energy level crossings of eigenstates of generic Hamiltonians \cite{sharmila2020signatures} and  their relation to quantum chaos \cite{karthik2007entanglement}, and the connection between entanglement and uncertainty relations \cite{paul2023tomographic}.
 
Coupled oscillator systems provide an ideal platform for investigating entanglement dynamics, 
and the roles played 
by  specific parameters and interactions (between different subsystems) on the nature of entanglement. 
Diverse aspects of entanglement have been examined using 
such systems both for obtaining exact  analytical results and for numerical studies on entanglement quantifiers. 
These investigations include (but are not confined to)  numerical studies of entanglement dynamics of 
two coupled oscillators with time-dependent frequencies, 
adopting specific methods using Gaussian states to estimate the Wigner function  \cite{dinc2021entanglement}, 
understanding the nature of the thermal state of 
such systems
when sudden changes of frequencies occur, leading to a thermal entanglement phase transition \cite{park2020thermal}, 
the entanglement between coherent states of the two oscillators \cite{jellal2011entanglement}, the role of quantum quenching on the entanglement measures \cite{choudhury2022entanglement}, the links between the dynamics of entanglement and the 
variations in  uncertainties of appropriate observables for both quenched and unquenched system of 
oscillators \cite{park2018dynamics}, and  studies on entanglement in systems 
with three oscillators \cite{park2019dynamics, merdaci2020entanglement}. Further, entanglement in such coupled oscillator models 
has been examined in a variety of field-theoretic contexts as well \cite{choudhury2023circuit,adhikari2023circuit, adhikari2023krylov, choudhury2022cosmological, nakagawa2016entanglement}.

Commonly used entanglement measures and monotones are defined through the density matrix of the system concerned. The density matrix itself has to be 
reconstructed from experimentally obtained tomograms which are histograms of measured observables. The state (i.e., density matrix) reconstruction program turns out to be a challenging task if the dimensionality of the Hilbert space is large, as in  CV and HQ systems and in certain DV systems.  
Hence, two commonly used  bipartite entanglement measures, the subsystem linear entropy (SLE) and the subsystem von Neumann entropy (SVNE) which are defined in terms of the density matrix, are not always easy to  compute. In terms of the reduced density matrix $\widetilde{\rho}$  of either subsystem of a bipartite system described by a pure state, the SLE is $1-\mathrm{Tr}\,(\widetilde{\rho}^{\,2})$, while the SVNE is given by $-\mathrm{Tr}\,(\widetilde{\rho}\ \mathrm{ln}\,\widetilde{\rho})$. Both these quantities vanish identically for unentangled states. In general, the Schmidt decomposition  (i.e., the diagonalization of $\widetilde{\rho}$) that is necessary for computing these measures becomes a  challenging numerical problem  for 3-dimensional CV systems. This is because a nontrivial integral equation has to be solved  to obtain the eigenvalues (or Schmidt coefficients)  
in a continuous basis.  Moreover, in certain cases the SLE and the SVNE may not be useful in quantifying entanglement. For instance,  even in the simple example of the electron-proton entanglement in the hydrogen atom, when the COM is delocalized, the SVNE diverges while  SLE tends to unity (its maximum value) \cite{mallayya2014zero, chandran2019divergence, qvarfort2020hydrogenic} for all energy eigenstates. Hence these measures are unsuitable for relating entanglement with the energy spectrum. Early investigations on the ground state of the hydrogen atom, with the center of mass (COM) delocalized  
\cite{tommasini1998hydrogen}, revealed that the spread in the Schmidt coefficients mirrors the spread in the momentum of the subsystem. The latter, which is easier to compute, could therefore be a possible indicator of entanglement. It has been shown \cite{qvarfort2020hydrogenic} that this spread scales inversely as the principal quantum number  for excited states. Further, if the COM is localized as a Gaussian wave packet, the second moments of the position and momentum could be used to identify entanglement.

The aim of the present investigation is to use tomographic entanglement indicators (TEIs) that are readily calculable from appropriate tomograms, to 
assess entanglement in coupled oscillator systems. This approach to the estimation of bipartite entanglement 
bypasses the challenging intermediate task of state reconstruction. In order to study the effects of factors such as the localization of the COM on entanglement, we analyze a 1-dimensional model of two coupled harmonic oscillators that is analytically tractable, and provides information on 
the performance of the TEIs.  Earlier literature on such bipartite oscillators \cite{srednicki1993entropy, makarov2018coupled, chandran2019divergence} primarily contains results pertaining to the  SVNE as a measure of entanglement between the two individual oscillators in any of the eigenstates of the composite 
system. These results corroborate the findings in the case of the hydrogen atom, namely,  that the SVNE diverges when the COM is not localized. 
Our emphasis is on the TEIs, and wherever necessary we will compare their performance with the SVNE and SLE.

The organization of the rest of this paper is as follows. In Sec. (\ref{section_tei_expln}) we define the TEIs that are used in this work. Section (\ref{section_ent_bw_ho}) deals with the description of the model of interest, the results obtained, and the inferences drawn. 
In  Sec. (\ref{section_concln}), we summarize our main results and suggest avenues for further research.

\section{Tomographic Entanglement Indicators}\label{section_tei_expln}
A quantum tomogram is a histogram comprising measurement outcomes of observables in different basis sets. For pure states of a given system, the tomogram 
of relevance to the present work is simply a collection of the expectation values of the density matrix in a quorum of basis sets. As a consequence, 
a tomogram does not explicitly involve the off-diagonal elements of the density operator. All the elements of the bipartite density matrix can be 
obtained, however, by carrying out an inverse Radon transformation on the tomogram. This is a consequence of the fact that, in general, the off-diagonal 
elements of the density operator are captured by performing the 
transformation on the collection of measurements in several basis sets \cite{man2013introduction}. In principle,
measurements in an infinite number of basis sets are required for this purpose \cite{mancini1997classical}. In practice, however, in the tomographic approach 
properties such as the bipartite entanglement in any system are quantified by computing the TEIs in a finite 
number of basis sets, and then evaluating their net contribution
using averaging procedures. In this context, it is worth recalling the following  question \cite{pauli} posed by 
Pauli in the early days of quantum mechanics: If 
the probability densities in the position and momentum basis alone are given, can the state of the system be uniquely 
reconstructed? The answer is that this is not generally possible (see, for instance, \cite{reichenbach}).  
As illustrated in the sequel, it turns out that in the model we consider the quorum is simply given by the 
position basis and momentum basis, and the averaging procedure yields results compatible with that of standard 
entanglement measures. However, this does not hold good in general.

Two aspects of the tomographic approach need to be kept in mind: first, 
the quorum is system-dependent, and no general prescription is available for identifying the number of basis 
sets for a given system. Second, the extent of entanglement depends on the basis selected for the measurement, 
and hence an averaging procedure is needed to extract the actual quantifier of entanglement. 

As already mentioned, reconstruction of the wave function of a many-body system is a 
computationally challenging task, even 
after suitable approximations are made. It is in this context that the TEIs could be very useful, if their slice-dependent 
(i.e., basis-dependent) signatures in different settings are examined in detail and categorized. 
Bearing these considerations in mind, we have carried out the computation of TEIs from specific slices of the tomogram for known states 
of the coupled oscillators. That is, we shall use the known position and momentum space oscillator probability distributions as marginals in the
calculation of the TEIs. The efficacy of the TEIs can then be judged by a comparison with the computed SVNE and SLE corresponding to the states 
concerned. In this sense, we are merely demonstrating the feasibility of the tomographic approach in this work.    

We recall that the tomogram is a collection of normalized probability distributions computed in different basis sets. A TEI is expected to assess the difference between a probability distribution corresponding to an entangled state and the factored product of the distributions  corresponding to the unentangled subsystem states. This exercise can be carried out readily if we express the TEIs in terms of `distances' between these distributions \cite{1089532}. Two examples of the latter are the Bhattacharyya distance ($\mathrm{\epsilon_{_{BD}}}$) and  the Kullback-Liebler divergence ($\mathrm{\epsilon_{_{KL}}}$), commonly used  in classical probability theory. 
While neither qualifies to be a distance in the strict sense of the term, they are useful as entanglement indicators.  We define them briefly below.

The Bhattacharyya distance quantifies the similarity between two probability distributions by estimating their 
mutual information. For a given slice of the tomogram, we have  
 \begin{eqnarray}\label{bd_defn}
            \mathrm{\epsilon_{BD}}=-\mathrm{log}_2 \; \Big\{\!\int \!\! d\bm{X} \! \int \!\! d\bm{Y}\big[
                \bra{\bm{X},\bm{Y}}\rho\ket{\bm{X},\bm{Y}} \bra{\bm{X}}\widetilde{\rho}_{1}\ket{\bm{X}}\bra{\bm{Y}}\widetilde{\rho}_{2}\ket{\bm{Y}}\big]^{1/2}\Big\},
        \end{eqnarray}       
where $\bm{X}$ and $\bm{Y}$ are the basis labels of  subsystems 1 and 2 respectively. In the example we will consider subsequently, these are the position (or momentum) 
labels of the two subsystems. $\rho$ is the full density matrix while $\widetilde{\rho}_{1}$, $\widetilde{\rho}_{2}$ are the reduced density matrices. 
The Kullback-Leibler divergence  is given by
\begin{eqnarray}\label{kl_defn}
        \mathrm{\epsilon_{KL}}=\int \!\! d\bm{X} \! \int \!\! d\bm{Y} \!\bra{\bm{X},\bm{Y}}\rho\ket{\bm{X},\bm{Y}}\mathrm{log}_2\; \frac{\bra{\bm{X},\bm{Y}}\rho\ket{\bm{X},\bm{Y}}}{\bra{\bm{X}}\widetilde{\rho}_{1}\ket{\bm{X}}\bra{\bm{Y}}\widetilde{\rho}_{2}\ket{\bm{Y}}}.
        \end{eqnarray} 
Both $\mathrm{\epsilon_{BD}}$ and $\mathrm{\epsilon_{_{KL}}}$ lie in the range [0, $\infty$), and vanish 
identically for unentangled states.
Another entanglement indicator $\mathrm{\epsilon_{_{IPR}}}$, based on the  inverse participation ratio (IPR),  can also be computed directly from the 
tomogram, as it involves elements of the density matrix in specific basis sets. The IPR was introduced originally in the context of estimating the localization properties of states. It indicates 
how `spread out' the state is in a given basis, and 
is related to a generalized definition of entanglement \cite{viola2007generalized}. By scaling the variables and the density matrix elements  
appropriately, we obtain the corresponding dimensionless quantities. Denoting the latter  with an overhead bar, the IPR-based entanglement indicator is 
given by 
\begin{equation}\label{ipr_defn}
        \mathrm{\epsilon_{_{IPR}}} = 1+\int \!\! d\overline{\bm{X}} \! \int \!\! d\overline{\bm{Y}}\bra{\overline{\bm{X}},\overline{\bm{Y}}}{\overline{\rho}}\ket{\overline{\bm{X}},\overline{\bm{Y}}}^2 -\int \!\! d\overline{\bm{X}}\bra{\overline{\bm{X}}}{\overline{\widetilde{\rho}}}_{1}\ket{\overline{\bm{X}}}^2 - \int \!\! d\overline{\bm{Y}}\bra{\overline{\bm{Y}}}{\overline{\widetilde{\rho}}}_{2}\ket{\overline{\bm{Y}}}^2.
\end{equation}
The precise transformation to the dimensionless quantities in Eq.~(\ref{ipr_defn}) depends on the model considered (Eqs. (\ref{dimensionless_var_rel_1}) and~(\ref{dimensionless_var_rel_2}) in the case of the model we study). 

\section{Entanglement between two harmonic oscillators}\label{section_ent_bw_ho}
Consider two coupled oscillators, each with mass $m$ and frequency $\omega$. The Hamiltonian is given by
 \begin{eqnarray}\label{ho_hamiltonian}
    H=\frac{1}{2m}({p_1}^{2}+{{p_2}^{2}}) +\frac{1}{2}m\omega^2 ({x_1}^{2}+{x_2}^{2})+\frac{1}{2}m\lambda{x_1}{x_2},
\end{eqnarray}
in the customary notation, $\lambda$ being the coupling constant between the oscillators. $H$ is positive definite if 
$\vert\lambda\vert < 2\omega^2$. In terms of the relative position $\displaystyle {x}={x_1}-{x_2}$,  the relative momentum $\displaystyle {p}=\tfrac{1}{2}({p_1}-{p_2})$, the COM coordinate $\displaystyle {X}=\tfrac{1}{2}({x_1}+{x_2})$ and the total momentum $\displaystyle {P}={p_1}+{p_2}$, $H$ has the separable form
\begin{eqnarray}\label{ho_hamiltonian_separable}
    H=\frac{{P}^{2}}{4m}+\frac{{p}^{2}}{m }+m\omega_c^2 {X}^{2}+\frac{m}{4} \omega_r^2 {x}^{2},
\end{eqnarray}
where $\omega_c^2=\omega^2+\frac{1}{2}\lambda$ and $\omega_r^2=\omega^2-\frac{1}{2}\lambda$.
It is convenient to work in a basis set of direct products $\ket{n_{c}, n_{r}} \equiv \ket{n_c} \otimes \ket{n_r}$ of the energy eigenkets of the COM and 
the relative system,  where $n_{c}$, $n_{r}$ $= 0,1,2,\dotsc$ .
In terms of the raising operators 
\begin{equation}
    a_r^{\dagger}=\Big(\frac{m\omega_r}{4\hbar}\Big)^{1/2}\Big({x}-\frac{2ip}{m\omega_r}\Big), \;\; a_c^{\dagger}=\Big(\frac{m\omega_c}{\hbar}\Big)^{1/2}\Big({X}-\frac{iP}{2m\omega_c}\Big)
\end{equation}
and the corresponding lowering operators, the Hamiltonian becomes   
\begin{equation}\label{ho_hamiltonian_ladder_rc}
    H=\hbar\omega_r\big(a_r^{\dagger}a_r+\tfrac{1}{2}\big)+\hbar\omega_c\big(a_c^{\dagger}a_c+\tfrac{1}{2}\big),
\end{equation}
whereas in terms of the ladder operators $a_i$ and $a_i^\dagger$ $(i = 1, 2)$ corresponding to the original pair of coupled oscillators, the Hamiltonian is
\begin{equation}\label{ho_hamiltonian_ladder_12}
    H=\hbar\omega(a_1^{\dagger}a_1+a_2^{\dagger}a_2+1)+\frac{\hbar\lambda}{\omega}(a_1^{\dagger}a_2^{\dagger}+a_1^{\dagger}a_2+a_1a_2^{\dagger}+a_1a_2).
\end{equation}  

\subsection{Uncoupled oscillators ($\mathbf{\lambda=0}$)}
 In the special case $\lambda=0$, the oscillators $1$ and $2$ are uncoupled and $\omega_c=\omega_r=\omega$.  The eigenstates of the Hamiltonian can then  be written in terms of the factored products of the energy eigenbasis of the two oscillators. For notational clarity, we will denote $\ket{n_1, n_2}$ by $\ket{\nu_1, \nu_2}$ and $\ket{n_c, n_r}$ by $\ket{\nu_c, \nu_r}$ in the case $\lambda = 0$. Further, we have in this case $\nu_1 + \nu_2 = \nu_c + \nu_r$, and
 $a_c^{\dagger}=(a_1^{\dagger}+a_2^{\dagger})/\sqrt{2}$, $a_r^\dagger = (a_1^{\dagger}-a_2^{\dagger})/\sqrt{2}$.

\begin{figure}[h]
\begin{center}
\includegraphics[width=0.5\textwidth]{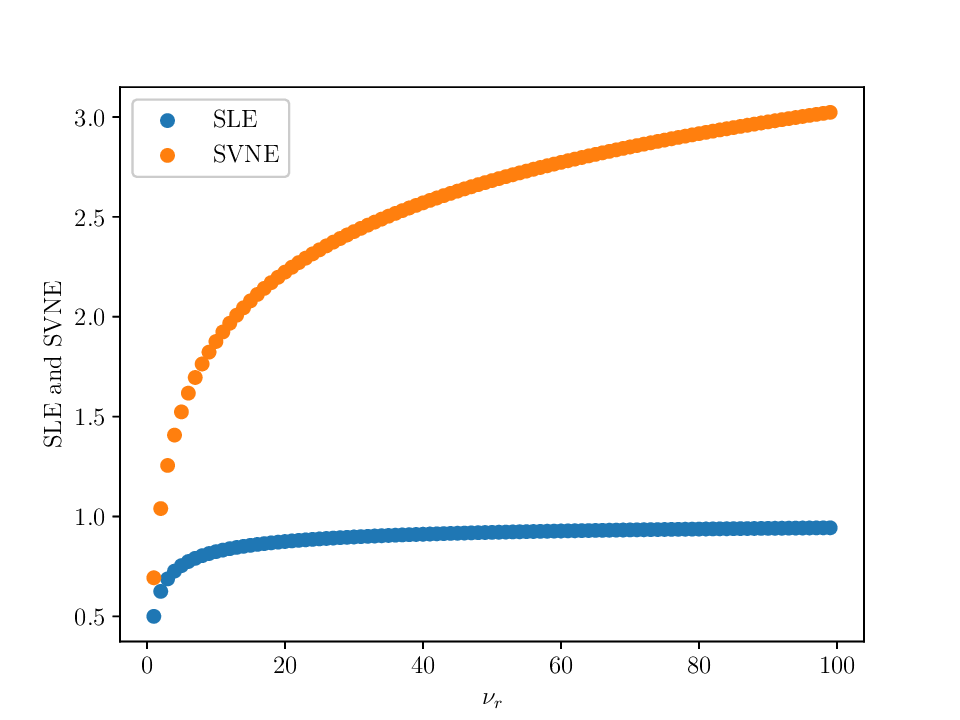}
\end{center}
\caption{SLE (Eq.~(\ref{eq:SLE_uncoup})) and SVNE (Eq.~(\ref{eq:SVNE_uncoup})) as a function of $\nu_r$ (with $\nu_c$=0) in the uncoupled case ($\lambda=0$).}
\label{sle_svne_uncoupled}
\end{figure}

We now consider the bipartite density matrix
\begin{align}
    \label{dens_mat_equal_omegas}
        \rho & =  \ket{\nu_c, \nu_r} \bra{\nu_c, \nu_r} \nonumber \\
        & = \sum_{\nu_1,\nu_2}\sum_{\nu_1',\nu_2'}\braket{\nu_1,\nu_2|\nu_c,\nu_r}\braket{\nu_c,\nu_r|\nu_1',\nu_2'}\ket{\nu_1, \nu_2} \bra{\nu_1', \nu_2'},
\end{align} 
where
\begin{equation}\label{schmidt_coeff_general_defn}
    \braket{\nu_1,\nu_2|\nu_c,\nu_r}= \frac{1}{(2^{\nu_c + \nu_r} \nu_c!\nu_r!)^{1/2}}\bra{\nu_1,\nu_2}{} (a_1^{\dagger}+a_2^{\dagger})^{\nu_c} (a_1^{\dagger}-a_2^{\dagger})^{\nu_r}\ket{\nu_c=0, \nu_r=0}.
\end{equation}
In explicit form, 
\begin{align}\label{schmidt_coeff_evaluation}
    \braket{\nu_1,\nu_2|\nu_c,\nu_r}= &\sum_{k=0}^{\nu_r}\sum_{k'=0}^{\nu_c} \Big\{\frac{(-1)^{k'}[\nu_c!\, \nu_r! \, (\nu_r+\nu_c-k-k')! \, (k+k')!]^{1/2}}{2^{(\nu_c+\nu_r)/2}(\nu_r-k)!\, k!\, (\nu_c-k')!\, k'!} \, \times \nonumber \\ & \delta_{\nu_1, \nu_r+\nu_c-k-k'}\, \delta_{\nu_2, k+k'} \Big\}.       
\end{align}
Tracing out oscillator 2 in Eq.~(\ref{dens_mat_equal_omegas}), the reduced density matrix becomes
\begin{align}\label{dens_mat_equal_omegas_n1n2_basis_step1}
    \widetilde{\rho} & = \sum_{\nu_1,\nu_1'}\braket{\nu_1,\nu_c+\nu_r-\nu_1|\nu_c,\nu_r} \delta_{\nu_1, \nu_1'} \braket{\nu_c,\nu_r|\nu_1',\nu_c+\nu_r-\nu_1'}\ket{\nu_1}\bra{\nu_1'} \nonumber \\
    & = \sum_{\nu_1}|\braket{\nu_1,\nu_c+\nu_r-\nu_1|\nu_c,\nu_r}|^2\ket{\nu_1}\bra{\nu_1}.
\end{align}
Since the Schmidt coefficients $|\braket{\nu_1,\nu_c+\nu_r-\nu_1|\nu_c,\nu_r}|^2$ are available explicitly from Eq.(\ref{schmidt_coeff_evaluation}), the SLE and  SVNE can be obtained analytically when $\lambda = 0$. In particular, in the case $\nu_c = 0$ these simplify to yield

\begin{equation}
\mathrm{SLE}=1-{2^{-2\nu_r}}\binom{2\nu_r}{ \nu_r},
\label{eq:SLE_uncoup}
\end{equation}


\begin{equation}
    \mathrm{SVNE}=\nu_r \ln 2- {2^{-\nu_r}}\sum_{l=0}^{\nu_r}\binom{\nu_r}{l} \ln \binom{\nu_r}{l}.
\label{eq:SVNE_uncoup}    
\end{equation}
Figure~\ref{sle_svne_uncoupled} shows the SLE and SVNE as functions of $\nu_r$ (when $\nu_c = 0$) in the uncoupled case $\lambda = 0$. We note that, in this instance, the entanglement implied by nonzero values of the SLE and SVNE is purely due to the state involved. The SLE saturates to unity as $\nu_r \to \infty$ with a leading asymptotic behavior $1 - (\pi \nu_r)^{-1/2}$, while the SVNE diverges in this limit.
\noindent In what follows we examine, in the general case $\lambda \neq 0$, both the standard entanglement measures as well as the TEIs, and compare the two sets of quantities.
\subsection{Coupled oscillators ($\mathbf{\lambda \neq 0}$): SLE and SVNE}
\begin{figure}[h]
\begin{center}
\begin{subfigure}{0.49\textwidth}
\includegraphics[width=\textwidth]{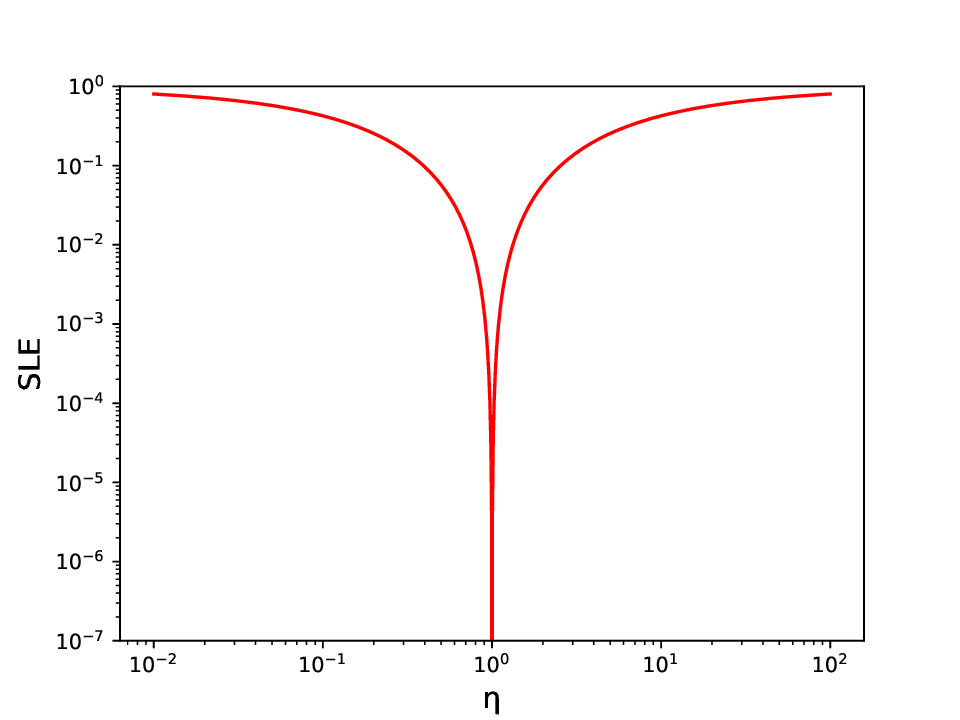}
\caption{} \label{sle0loglog}
\end{subfigure} 
\begin{subfigure}{0.49\textwidth}
\includegraphics[width=\textwidth]{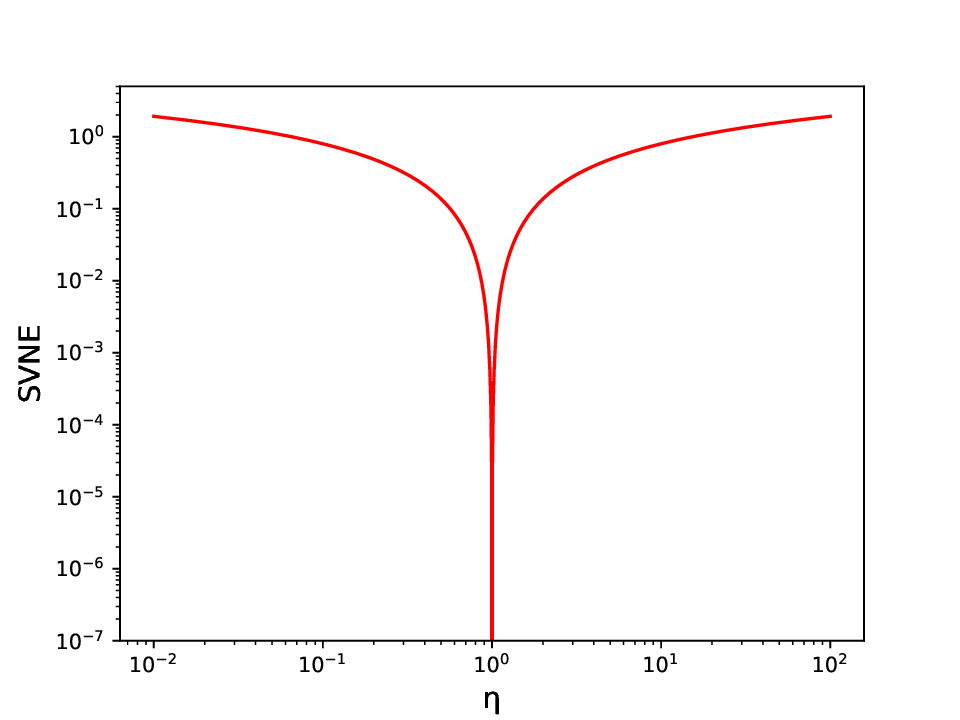}
\caption{}\label{svne0loglog}
\end{subfigure}
\end{center}
\caption{Log-log plots of the SLE (Eq.(\ref{sle_ground_state})) and SVNE (Eq.(\ref{svne_ground_state})) versus $\eta$ in the case $n_c = 0$, $n_r = 0$.}
\end{figure}

We analyze two cases in detail, with the COM oscillator in its ground state ($n_c= 0$) in both instances: Case (a) 
$n_r=0$, and  Case (b) $n_r \geqslant 1$. We first compute the SLE and SVNE in Case (a), for which both these 
entropies can be found analytically. Noting that the COM and the relative coordinate wave functions are Gaussians 
in $X$ and $x$ respectively, the matrix elements of the density
matrix in the position basis of the oscillators 1 and 2 work out to
\begin{equation}\label{dens_mat_ground_state_12}
    \braket{x_1,x_2|\rho|x_1', x_2'} = (4/\pi)\, \sqrt{c_1 c_2}\,e^{-c_1 [({x_1+x_2})^2+(x_1'+x_2')^2]} e^{-c_2 [(x_1-x_2)^2+(x_1' - x_2')^2]},
\end{equation}
where $c_1=m\omega_c/4\hbar$ and $c_2=m\omega_r/4\hbar$. Tracing over oscillator $2$, the reduced density matrix element becomes
\begin{equation}\label{red_dens_mat_ground_state_12}
    \braket{x_1|\widetilde{\rho}|x_1'} =\Big[\frac{8c_1 c_2}{\pi(c_1+c_2)}\Big]^{1/2} \, \mathrm{exp} \, \Big\{-(c_1+c_2)(x_1^2+x_1'^2) + \frac{(c_1-c_2)^2}{2(c_1+c_2)}(x_1+x_1')^2\Big\}.
\end{equation}
The SLE can now be computed in a straightforward manner. In terms of the ratio $\eta=\omega_c/\omega_r$, it is given by
\begin{equation}\label{sle_ground_state}
        1 - \mathrm{Tr}\, \widetilde{\rho}^{\,2} = \frac{(1-\sqrt{\eta})^2}{1+\eta}.
\end{equation}
As expected, the SLE vanishes when $\eta = 1$. We also note that it is unchanged under $\eta \to 1/\eta$, i.e., the interchange 
$\omega_c \leftrightarrow \omega_r$ (see Fig.\ref{sle0loglog}).

A similar procedure cannot be used to compute the SVNE because the matrix elements of $\mathrm{ln}\ \widetilde{\rho}$ in the position basis are not 
readily calculable analytically. However, for the case $n_c=n_r=0$ alone (with which we are presently concerned), an analytic expression for 
the Schmidt coefficients in terms of the system parameters can be obtained starting from a trial ansatz \cite{srednicki1993entropy}. Further, 
a mapping between $\widetilde{\rho}$ and the density matrix of a single oscillator in a heat bath has also been mentioned therein. We 
indicate below the computation of the SVNE using such a mapping. The reduced density matrix of a single oscillator of frequency $\varpi$ in a 
heat bath at inverse temperature $\beta$ is given by $\widetilde{\rho}=e^{-\beta {\mathcal{H}}}/{\mathrm{Tr}\,e^{-\beta {\mathcal{H}}}}$, where $\mathcal{H}$ is its Hamiltonian.
The eigenvalues of $\widetilde{\rho}$ are given by
\begin{eqnarray}\label{eigenvalues_single_osc_heat_bath}
 \lambda_n=(1-e^{-\beta \hbar \varpi})\,e^{-n\beta\hbar\varpi},  \quad n = 0, 1, 2, \dotsc
\end{eqnarray}
It follows at once that the SLE is
\begin{equation}
1 - \mathrm{Tr} \, \widetilde{\rho}^{\, 2} = 1-\mathrm{tanh}\, \tfrac{1}{2} \beta \hbar \varpi,
\end{equation}
while the SVNE is
\begin{equation}
\label{eqn:SVNE}
 - \mathrm{Tr} \, (\widetilde{\rho}\,\, \mathrm{ln}\, \widetilde{\rho}) = -\mathrm{ln}\,(1-e^{-\beta \hbar \varpi}) + \frac{\beta \hbar \varpi}{e^{\beta \hbar \varpi}-1}.
\end{equation}
The matrix elements of $\widetilde{\rho}$ in the position basis are given by
\begin{equation}\label{dens_mat_single_osc_heat_bath_position_basis}
    \braket{x_1|\widetilde{\rho}|x_1'}=\sum_{n = 0}^\infty \lambda_n\, \chi_n(x_1)\, \chi_n(x_1'),
\end{equation}
where $\chi_n$ is the oscillator wave function given by
\begin{equation}
    \chi_n(x_1)= \frac{(m\varpi/\pi \hbar)^{1/4}}{(2^n n!)^{1/2}}H_n(x_1\sqrt{m\varpi/\hbar})\, e^{- m\varpi x_1^2/2\hbar}
\end{equation}
and $H_n$ is the Hermite polynomial of order $n$. The sum in Eq. (\ref{dens_mat_single_osc_heat_bath_position_basis}) is evaluated using Mehler's formula \cite{mehler}
\begin{equation}\label{Mehler_formula}
    \sum_{n = 0}^{\infty}H_n(u)H_n(v)\frac{s^n}{2^n n!}=(1-s^2)^{-1/2} \,\mathrm{exp}\,\Big\{-\frac{s^2u^2+s^2v^2-2uvs}{1-s^2}\Big\} , \ |s|<1.
\end{equation}
The result is
\begin{align}\label{dens_mat_single_osc_heat_bath_position_basis_evaluated}
    \braket{x_1|\widetilde{\rho}|x_1'}  =\Big[\frac{m\varpi \, \mathrm{tanh}\, \tfrac{1}{2} \beta \hbar \varpi}{\pi \hbar}\Big]^{1/2} \mathrm{exp}\, \Big[-\frac{m\varpi}{2 \hbar}&\Big\{(x_1^2+x_1'^2)\,\mathrm{coth}\,\beta \hbar \varpi  \nonumber \\
    &- 2\,x_1\, x_1'\,\mathrm{cosech} \, \beta \hbar \varpi \Big\} \Big].
\end{align}
The expressions in Eqs. (\ref{dens_mat_single_osc_heat_bath_position_basis_evaluated}) and (\ref{red_dens_mat_ground_state_12}) are the same if we make the identifications 
 \begin{equation}\label{heat_bath_param_rel_om}
    \varpi=(\omega_c\,\omega_r)^{1/2}
\end{equation}
for the frequency of the oscillator, and
\begin{equation}\label{heat_bath_param_rel_beta}
    \beta=\frac{1}{\hbar \varpi}\, \displaystyle{\mathrm{ln}\,\Big(\frac{1+\sqrt{\eta}}{1-\sqrt{\eta}}\Big)^2}
\end{equation}   
for the inverse temperature of the heat bath.
Since we already have the expressions for the SLE and SVNE for this system, we can now readily identify the corresponding 
entanglement measures for the original problem of the two coupled oscillators. Equation (\ref{sle_ground_state}) for the SLE is immediately 
corroborated. For the SVNE, inserting the correspondences of Eqs.~(\ref{heat_bath_param_rel_om}) and~(\ref{heat_bath_param_rel_beta}) in Eq.~(\ref{eqn:SVNE}) we get
\begin{equation}\label{svne_ground_state}
    - \mathrm{Tr} \, (\widetilde{\rho}\,\, \mathrm{ln}\, \widetilde{\rho}) =-\mathrm{ln}\,\Big(\frac{4\sqrt{\eta}}{(1+\sqrt{\eta})^2}\Big) 
                 +\frac{(1-\sqrt{\eta})^2}{4\sqrt{\eta}}\mathrm{ln}\,\Big(\frac{1+\sqrt{\eta}}{1-\sqrt{\eta}}\Big)^2.
\end{equation}

\noindent The SVNE diverges as $\eta \to 0$ (or $\omega_c \to 0$), since the COM oscillator is delocalized \cite{qvarfort2020hydrogenic} in this case. It vanishes when 
$\eta = 1$, as expected, and is symmetric (like the SLE, Eq.~(\ref{sle_ground_state})) under the interchange $\eta \leftrightarrow 1/\eta$. 
This symmetry is manifest in the log-log plot of the SVNE versus $\eta$ as shown in Figs.~\ref{sle0loglog} and ~\ref{svne0loglog}.

\begin{figure}
\begin{subfigure}{0.49\textwidth}
\includegraphics[width=\textwidth]{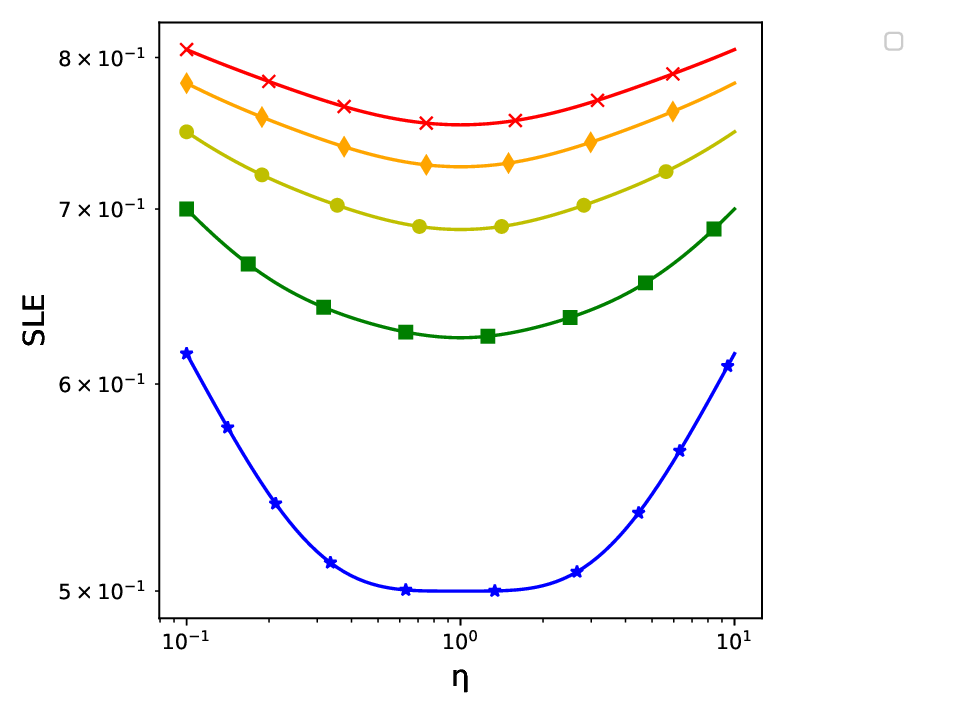}
\caption{} \label{sleone}
\end{subfigure} 
\begin{subfigure}{0.49\textwidth}
\includegraphics[width=\textwidth]{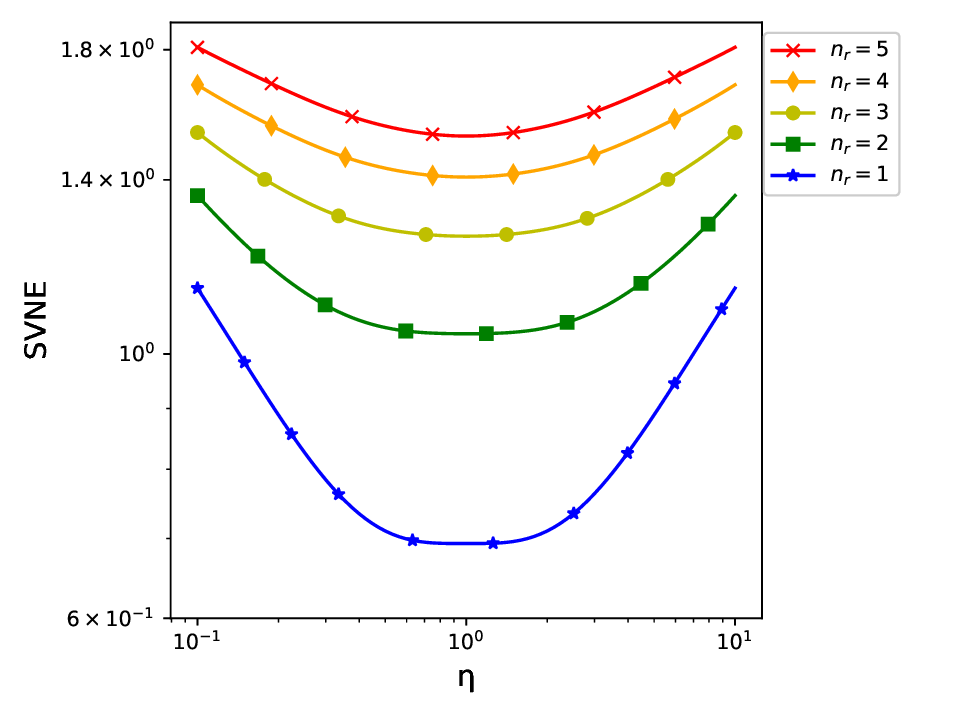}
\caption{}\label{svneone}
\end{subfigure}
\caption{Log-log plots of the (a) SLE and (b) SVNE for $n_c=0$, $n_r\geqslant 1$.}\label{results}
\end{figure}

We now proceed to discuss Case (b), i.e., $n_c=0$, $n_r \geqslant 1$. 
The Schmidt coefficients, and hence the extent of entanglement between the coupled oscillators 1 and 2,  have 
been obtained as functions of $\eta$ by numerically diagonalizing the reduced density matrix. Plots of the SVNE and SLE versus $\eta$, for a 
set of values of $n_r$, are shown in Figs.~\ref{sleone} and~\ref{svneone}. It is clear that both these entanglement measures increase with 
$n_r$, or equivalently, with the energy of the oscillator in the relative coordinate. 

A general remark is in order at this juncture. The entanglement in the system under study arises from two sources: (i) the 
choice of the state of the system, and (ii) the effect of a non-zero coupling ($\lambda \neq 0$). The first of these is 
demonstrated by Eqs.~(\ref{eq:SLE_uncoup}),~(\ref{eq:SVNE_uncoup}) and Fig.~\ref{sle_svne_uncoupled} showing the entanglement
due to the state ($\nu_c = 0, \nu_r \geqslant 1$) in the absence of coupling. In the second case, on the other hand, Eqs.~(\ref{sle_ground_state}) 
and~(\ref{svne_ground_state}) show that entanglement occurs even in the ground state $n_c = 0, n_r = 0$ owing to a non-vanishing coupling ($\lambda \neq 0$). In general the contribution to entanglement from the choice of state can be identified
by passing to the limit $\lambda = 0$ in the expressions for the SLE and SVNE in each case.

\subsection{Coupled oscillators: Tomographic entanglement indicators}
\begin{figure}[h]
    \begin{center}
    \includegraphics[width=\textwidth]{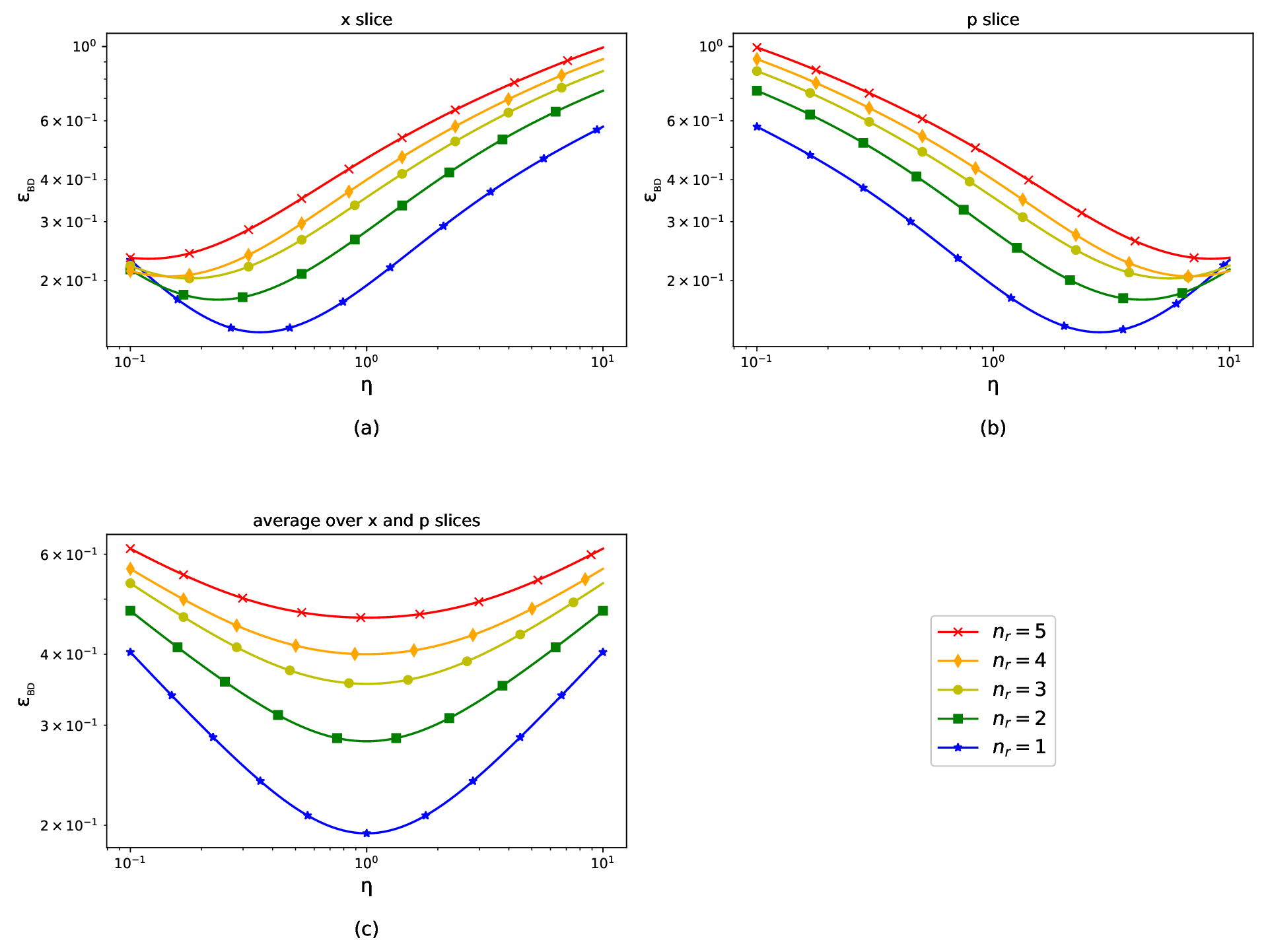}
    \end{center}
    \caption{Log-log plot of $\mathrm{\epsilon_{_{BD}}}$ versus $\eta$ for $n_c = 0$, $n_r \geqslant 1$. Top left and right, position and momentum slices respectively; bottom, average of position and momentum slices.}
    \label{fig:eps_bd_osc}
\end{figure}

\begin{figure}[h]
\begin{center}
\includegraphics[width=0.6\textwidth]{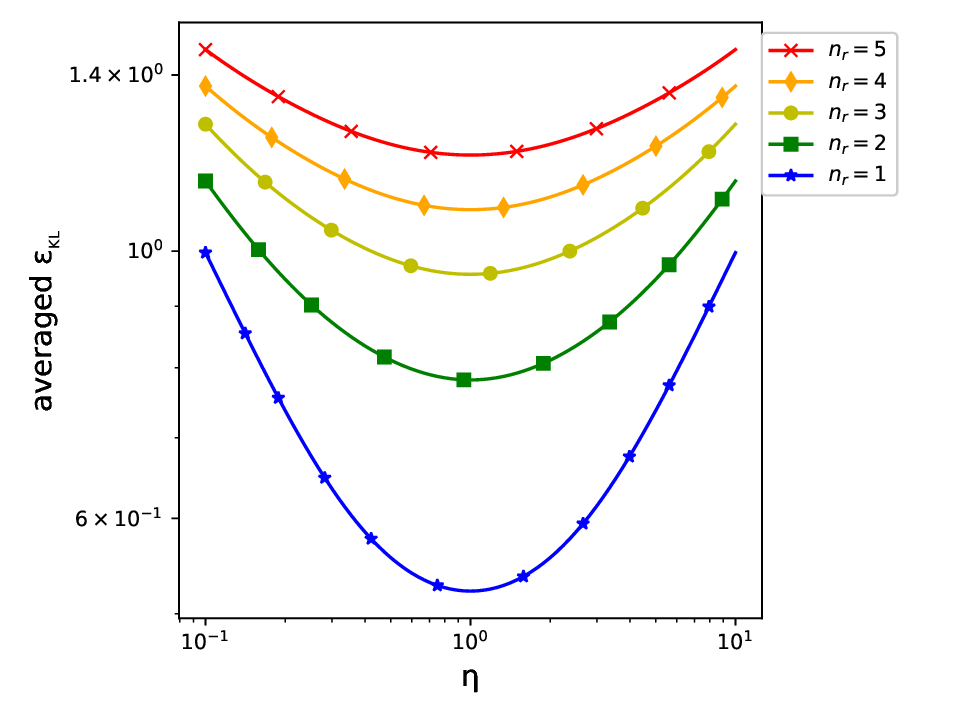}
\end{center}
\caption{Log-log plot of $\mathrm{\epsilon_{_{KL}}}$ averaged over the position and momentum slices versus $\eta$, for $n_c = 0$ and $n_r \geqslant 1$.}
\label{fig:eps_KL_osc}
\end{figure}

We now calculate the TEIs (defined in Eqs.~(\ref{bd_defn})--(\ref{ipr_defn})) in our oscillator model.
Once again, in the limit $\eta\rightarrow0$ it can be shown that both $\mathrm{\epsilon_{_{BD}}}$ and 
$\mathrm{\epsilon_{_{KL}}}$ diverge as a consequence of the delocalization of the COM oscillator.  
Setting $n_c=0$ as before, plots of $\mathrm{\epsilon_{_{BD}}}$ versus $\eta$ for a range of values of $n_r$ 
have been obtained numerically and displayed in Fig. \ref{fig:eps_bd_osc}. We see that the minimum in $\mathrm{\epsilon_{_{BD}}}$ as a 
function of $\eta$ in Fig. \ref{fig:eps_bd_osc}a drifts to smaller values of $\eta$ with increasing $n_r$. This is an artefact arising
from our choice of the position slice of the tomogram. Figure \ref{fig:eps_bd_osc}b corresponds to the momentum slice of the same tomogram,  
in which the minimum drifts to larger values of $\eta$ with increasing $n_r$.
Figure \ref{fig:eps_bd_osc}c depicts the average of the values obtained from the two slices, showing that the minimum value of $\mathrm{\epsilon_{_{BD}}}$ remains at 
$\eta = 1$ independent of $n_r$, in accordance with the behavior of the SLE and SVNE. The same property is exhibited by the
averaged TEI $\mathrm{\epsilon_{_{KL}}}$, as seen in Fig. \ref{fig:eps_KL_osc}. Overall, the TEIs $\mathrm{\epsilon_{_{BD}}}$ and $\mathrm{\epsilon_{_{KL}}}$ do reflect quite faithfully the variations in the SLE and SVNE. 

\begin{figure}
\begin{center}
\includegraphics[width=0.5\textwidth]{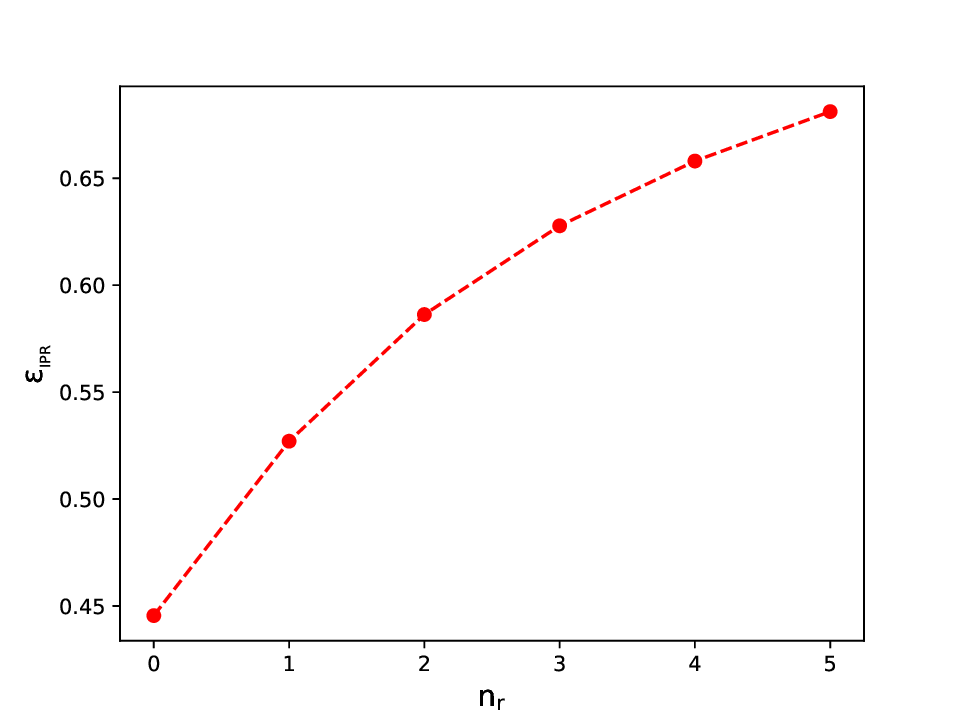}
\end{center}
\caption{$\mathrm{\epsilon_{_{IPR}}}$ (in the position slice) versus $n_r$, with $n_c = 0$ and $\eta = \tfrac{1}{4}$.}
\label{compare}
\end{figure}

We now proceed to compute $\mathrm{\epsilon_{_{IPR}}}$ (Eq. (\ref{ipr_defn})). We introduce 
dimensionless variables $\overline{x}= x/L_r$ and $\overline{X}= X/L_c$ where $L_r = (2\hbar/m\omega_r)^{1/2}$ and 
$L_c = (\hbar/2m\omega_c)^{1/2}$. Defining $\overline{x}_{1,2} = \overline{X} \pm \tfrac{1}{2}\overline{x}$ we have, in terms of the original 
oscillator coordinates $x_1$ and $x_2$,
\begin{equation}
    \label{dimensionless_var_rel_1}
        \overline{x}_1 = \tfrac{1}{2}(L_c^{-1} + L_r^{-1})x_1 + \tfrac{1}{2}(L_c^{-1} - L_r^{-1})x_2
\end{equation}
    and
\begin{equation}
    \label{dimensionless_var_rel_2}
        \overline{x}_2 = \tfrac{1}{2}(L_c^{-1} - L_r^{-1})x_1 + \tfrac{1}{2}(L_c^{-1} + L_r^{-1})x_2.
\end{equation}    
The further requirement that $\overline{x}_1$ and $\overline{x}_2$ be proportional to $x_1$ and $x_2$, respectively, imposes the condition 
$L_c=L_r$ ($=L$, say). This restricts $\eta$ to the value $\frac{1}{4}$. The dimensionless density matrix elements are given by
\begin{eqnarray}\label{dens_mat_dimensionless}
    \bra{\overline{x}_1,\overline{x}_2}{\overline{\rho}}\ket{\overline{x}_1',\overline{x}_2'}=L^2\bra{L x_1,L x_2}{\rho}\ket{L x_1',L x_2'}
\end{eqnarray}
and the corresponding TEI is given by
\begin{align}\label{ipr_oscillator}
    \mathrm{\epsilon_{_{IPR}}}=1+\int\!\! d\overline{x}_1 \!\!\int\!\! d\overline{x}_2\bra{\overline{x}_1,\overline{x}_2}{\overline{\rho}}\ket{\overline{x}_1,\overline{x}_2}^2 -\int\!\! d\overline{x}_1\bra{\overline{x}_1}{\overline{\widetilde{\rho}}}_{1}\ket{\overline{x}_1}^2-\int\!\! d\overline{x}_2\bra{\overline{x}_2}{\overline{\widetilde{\rho}}}_{2}\ket{\overline{x}_2}^2.
\end{align}
\noindent A numerically generated plot of $\mathrm{\epsilon_{_{IPR}}}$ versus $n_r$, in the position slice, for $\eta=\frac{1}{4}$ and $n_c=0$, is shown in 
Fig. \ref{compare}. It is again evident that $\mathrm{\epsilon_{_{IPR}}}$ increases with increasing $n_r$, following the same general trend as 
all the other indicators considered. 

\section{Concluding remarks}\label{section_concln}
We have investigated the entanglement between two coupled harmonic oscillators with the same natural frequency. The system is separable in 
the center of mass (COM) and relative coordinates. The standard entanglement measures (SLE and SVNE) have been found explicitly when the
oscillators in the COM and relative coordinates are in the ground state ($n_c = n_r = 0$), and numerically for $n_c = 0, n_r \geqslant 1$.  
In the special case when the oscillators in the COM and relative coordinates have the same frequency, Schmidt diagonalization has been  
carried out to yield the SLE and SVNE analytically. Tomographic entanglement indicators (the Bhattacharya distance, the Kullback-Liebler 
divergence, and the IPR-based indicator) have been computed. They have been shown to compare very favourably with the standard entanglement measures.
The TEIs also indicate that localization is crucial to ensure meaningful estimates of 
entanglement. (When the COM is not localized, all indicators reach their maximum values for all eigenstates.)  We have also shown that 
the entanglement increases with the energy of the oscillator in the relative coordinate. Further, we have 
identified and commented on the contributions to the entanglement from the initial state and the coupling between modes, 
respectively.

While the entanglement in coupled oscillator systems has been investigated in considerable detail in the 
literature, and extension to field theoretic models of such oscillator-based studies has been attempted, 
it is evident that with an increase in the number of oscillator modes, the construction of the net density 
operator, and hence the computation of the entanglement indicators, become difficult tasks. Our investigations indicate that the tomographic approach is a viable  alternative to procedures based on detailed state reconstruction. While 
bipartite entanglement studies alone have been attempted here, the encouraging results suggest that 
it is worth extending the tomographic method to systems with a larger number of interacting oscillators and then to other
multipartite systems.

It would also be desirable to extend the preliminary understanding of the effects of 
localization and binding on entanglement gained from the present work to three-dimensional systems and many-body CV systems in both atomic and nuclear 
physics. For instance, the roles played by the entanglement between the
spin and spatial wave functions, and that between the space-spin
state and the isospin state in determining the structure of a nucleus, merit detailed
attention. These aspects are under investigation.

\section*{Acknowledgements}
We are pleased to dedicate this work to Prof. A. R. P. Rau with whom we have had many fruitful 
discussions over the years.
We acknowledge partial support through funds from Mphasis to the Center for Quantum 
Information, Computing and Communication (CQuICC), IIT Madras. SL and VB thank the Department of Physics, Indian 
Institute of Technology Madras for infrastructural support.

\section*{Competing interests}
The authors declare there are no competing interests.

\section*{Data availability statement}
This manuscript does not report data.



\end{document}